\documentclass[a4paper]{article}

\usepackage[english]{babel}
\usepackage[utf8x]{inputenc}
\usepackage[T1]{fontenc}
\usepackage{amssymb}

\usepackage{amsmath}
\usepackage{graphicx}
\usepackage[colorinlistoftodos]{todonotes}
\usepackage[colorlinks=true, allcolors=blue]{hyperref}
\usepackage[affil-it]{authblk}

\title{Two  high capacity text steganography schemes  based on color  coding}
\author{Juvet K. Sadi\'e$^{1,2,3,4}$, Leonel Moyou Metcheka$^{1,2,3,4}$, Ren\'{e} Ndoundam$^{1,2,3,4,}$\footnote{Corresponding author: ndoundam@yahoo.com}\\ 
{\small $^1$Team GRIMCAPE} \\
{\small $^2$Sorbonne Unversity, IRD, UMMISCO, F-93143, Bondy, France} \\
{\small $^3$CETIC, Yaounde, Cameroon}\\
{\small $^4$Department of Computer Science, University of Yaounde I, P.o. Box 812 Yaounde, Cameroon}\\
{\small E.mail: sadie.juvet@gmail.com, leonelmoyou@gmail.com, ndoundam@yahoo.com}  
 }

\date{}
\begin{document}
\maketitle

\begin{abstract}
Text steganography is a mechanism of hiding secret text message inside another text as a covering message. In this paper, we propose a text steganographic scheme based on color coding. This include two different methods: the first based on permutation, and the second based on numeration systems. Given a secret message and a cover text, the proposed schemes embed the secret message in the cover text by making it colored. The stego-text is then sent to the receiver by mail. After experiments, the results obtained show that our models performs a better hiding process in terms of hiding capacity as compared to the scheme of Aruna Malik et al. in which our idea is based.
\end{abstract}
\pagebreak 

\section{Introduction}

The word steganography is of Greek origin and means covered writing.  It is the hiding of a message within another (cover medium) such as web pages, images or text, so that the presence of the hidden message is indiscernible.  When a message is hidden in the cover medium, the resulting medium is called a stego-object. The key concept behind steganography is that the message to be transmitted should not be detectable with bare eyes. From the definition, Steganography is used to ensure data confidentiality, like encryption. However, the main difference between the two method is that with encryption, anybody can see that both parties are communicating in secret. Steganography hides the existence of a secret message and in the best case nobody can detect the presence of the message. When combined, steganography and encryption can provide more security. Steganography dates back to ancient Greece, where common practices consisted of etching messages in wooden tablets and covering them with wax. A number of steganographic methods have been introduced on different cover media such as images \cite{t1, t2, t3}, video files \cite{t4, t5} and audio files \cite{t6}. In text based steganographic methods, text is used as a cover media for hiding the secret data. Due to the lack of large scale redundancy of information in a text file, the human eye is very susceptible to any change between the original and the modified texts. Therefore, text steganography seems to be the most difficult kind of steganography \cite{t7}, as compare to others.\\In this paper, we propose a text steganographic scheme based on color coding, permutation and numeration systems. Given a secret message and a cover text, the proposed scheme embed the secret message in the cover text by making it colored, using a permutation algorithm for the first method and numeration systems for the second one.
After the first section devoted to the introduction, section 2 presents some preliminaries and related works. Section 3 concerns the presentation of the first method of our scheme. Section 4 labels the second approach of our scheme, and finally conclusion is stated in section 5.

\section{Preliminaries and related works}
In this section, the focus is to present some preliminaries that lead us to the comprehension of our scheme. Also, we present related works in the field of text steganography.

\subsection{Text Steganography}
There are many techniques in text steganography. In Syntactical steganography, punctuation marks such as full stop (.), comma (,) etc, are used to hide bits in cover text. The problem with this method is that it requires identification of correct places to insert punctuation \cite{t8, t9}. In lexical steganography, words are used to hide secret bits. A word could be replaced by its synonyms and the choice of word to be chosen from the list of synonyms would depend on secret bits. Sms texting is a combination of abbreviated words used in sms \cite{t10}. This technique proposes to hide binary data by using full form or its abbreviated form. For instance, to hide 0, full form of the word is used and to hide 1, abbreviated form of word is used \cite{t10}. The CSS technique encrypts a message using RSA public key cryptosystem and cipher text is then embedded in a cascading style  Sheet (CSS) by using End of Line on each CSS style properties, exactly after a semi-colon. A space after a semi-colon embeds bit 0 and a tab after a semicolon embeds bit 1 \cite{t11}.
Anandaprova Majumder and al \cite{t12} proposed an approach for text steganography through a technique that uses reflection symmetry of the English alphabet. Ekodeck and Ndoundam \cite{t13} proposed different approaches of PDF file based steganography, essentially based on the Chinese Remainder Theorem. Here, after a cover PDF document has been released from unnecessary characters of ASCII code A0, a secret message is hidden in it using one of the proposed approaches, making it invisible to common PDF readers, and the file is then transmitted through a non-secure communication channel.
Aruna Malik and al \cite{t14}, proposed a high capacity text steganography scheme based on LZW compression and color coding. Their scheme uses the forward mail platform to hide secret data. The algorithm first compresses secret data and then hide the compressed data into the email addresses and also, in the cover message of email. The secret data is embedded in the message by making it colored using a color table. Here below, some limits of that scheme will be presented.

\subsection{Critic and limits}

LZW is a lossless compression technique that performs high compression ratio when the source contains repetition pattern. In the LZW based steganographic scheme propose by Aruna Malik \cite{t14}, they apply this lossless compression on the secret message to increase the embedding capacity. But in the example proposed, there is no compression. In other words, the size of the compressed text is much greater than the size of the secret. To show this, we will give three different implementations of LZW algorithm applied to the secret message.

\subsubsection{The LZW Algorithm with initial dictionary fixed and known} \label{subsubsection 2.2}
This algorithm \cite{t15} starts by initializing the dictionary with the 256 characters of the ASCII code from 0 to 255. The output codes start at a minimum bit size equal to 9 and in general, as long as the indexes considered are strictly inferior to n = $2^k$ - 1, we can represent them on k bits. When the first integer greater than or equal to $2^k$ - 1 is met, the sequence 1. . . 1 (k times bit 1) and continue with coding the integers on k + 1 bits. Applying this method to the following secret message: "underlying physiological mechanisms", we obtain the outputs presented in Table \ref{table1}. The binary compressed text is obtained by converting the indexes of the output column of the array to 9 bits : \\ 
001110101 001101110 001100100 001100101 001110010 001101100 001111001 001101001 001101110 001100111 000100000 001110000 001101000 001111001 001110011 001101001 001101111 001101100 001101111 001100111 001101001 001100011 001100001 001101100 000100000 001101101 001100101 001100011 001101000 001100001 001101110 001101001 001110011 001101101 001110011. Hence, the size of the output is 35*9 = 315 bits.

\begin{table}[t]
\centering
\begin{tabular}{|cccl|cccl|}
\hline 
Buffer & input-char & Output & New Item &	Buffer & input-char & Output & New Item\\
\hline
u      &    n    &    117    &   256=un &	l      &    o    &    108    &   273=lo \\
n      &    d    &    110    &   257=nd &	o      &    g    &    111    &   274=og \\
d      &    e    &    100    &   258=de &	g      &    i    &    103    &   275=gi \\
e      &    r    &    101    &   259=er &	i      &    c    &    105    &   276=ic \\
r      &    l    &    114    &   260=rl &	c      &    a    &    99     &   277=ca \\
l      &    y    &    108    &   261=ly &	a      &    l    &    97     &   278=al \\
y      &    i    &    121    &   262=yi &	l      &          &    108   &  279=l \\
i      &    n    &    105    &   263=in &	       &    m    &    32     &   280= m \\
n      &    g    &    110    &   264=ng &	m      &    e    &    109    &   281=me\\
g      &         &    103    &   265=g 	&	e      &    c    &    101    &   282=ec \\
       &    p    &    32     &   266= p &	c      &    h    &    99     &   283=ch \\
p      &    h    &    112    &   267=ph &	h      &    a    &    104    &   284=ha \\
h      &    y    &    104    &   268=hy &	a      &    n    &    97     &   285=an \\
y      &    s    &    121    &   269=ys &	n      &    i    &    110    &   286=ni \\
s      &    i    &    115    &   270=si &	i      &    s    &    105    &   287=is \\
i      &    o    &    105    &   271=io &	s      &    m    &    115    &   288=sm\\
o      &    l    &    111    &   272=ol &	m      &    s    &    109    &   289=ms \\
	   &		 &			 &			&	s	   &        &     115    &       \\
\hline  			
\end{tabular}
\caption{LZW Algorithm output with initial dictionary fixed and known}
\label{table1}
\end{table}

\subsubsection{The LZW algorithm with sharing of the initial dictionary} \label{subsubsection1 2.2}

In this version \cite{t15}, initial dictionary contains only the character of the secret message. The output code is represented on height bits. The particularity of this implementation is from the initial dictionary which must be shared between the two parties in order to be able to decompress the binary code. Table \ref{table2} presents the initial dictionary for the same secret message:
Here are the output code : 1  2  3  4  5  6  7  8  2  9 10 11 12  7 13  8 14  6 14  9  8 15 16  6 10 17  4 15 12 16  2  8 13 17 13 and in binary we have : \\
00000001 00000010 00000011 00000100 00000101 00000110 00000111 00001000 00000010 00001001 00001010 00001011 00001100 00000111 00001101 00001000 00001110 00000110 00001110 00001001 00001000 00001111 00010000 00000110 00001010 00010001 00000100 00001111 00001100 00010000 00000010 00001000 00001101 00010001 00001101.
Hence, the size of the output is the sum of the size of initial dictionary and the output code: 17+35 = 52 bytes = 416 bits.

\begin{table}\centering
\begin{tabular}{cccccccccccccccccc}
\hline 
Index & 1 &	2&3	&4	&5	&6	&7	&8	&9	&10	&11	&12	&13	&14	& 15&16 &17   \\
char  & u &	n&d	&e	&r	&l	&y	&i	&g	&space	&p	&h	&s	&o	&c &a & m \\
\hline  
  \end{tabular}
  \caption{ Initial Dictionnary}
    \label{table2}
\end{table}

\subsubsection{The Unix compress command} \label{subsubsection2 2.2}
The ncompress package \cite{t17} is a compression utility available on Linux which contains the compress command for fast compression and decompression using LZW algorithm. The algorithm behind this command is explained at page 153 of the data compression book \cite{t15}. The initial dictionary size is 512 and the minimum output code size is 9 bits. This package can be installed by using the command "sudo apt-get install ncompress". Based on the Ubuntu 16.04 platform, this command produces a file with .Z extension as compress file. By Applying this command "compress -v source.txt" to the secret message contained in text source file with the -v option, the given output indicates that there is no compression and the .Z file size is 44 bytes = 352 bits.\\
Finally the table  \ref{table3} shows the comparison in bit between the original text size and the output size after the compression using the three different approaches of LZW implementation: The LZW Algorithm with initial dictionary fixed and known, The LZW algorithm with sharing of the initial dictionary and The Unix compress command.
\\
\begin{table}\centering
\begin{tabular}{cccc}
\hline 
Secret message size & Output size 1 & Output size 2   & Output size 3 \\
\hline
   280   &   315   & 416 & 352    \\   
\hline
  \end{tabular}
   \caption{Secret message size comparison}
   \label{table3}
\end{table}

From Aruna and al. paper \cite{t14}, the size obtained was 264 bits, but we have proven above that there is no compression for this example. This is the principal limit of this steganographic scheme, where for some messages the reduction of the message size will not be possible. 
Our paper uses:
\begin{itemize}
	\item The idea of color coding contained in the paper of Aruna Malik and al \cite{t14};
	\item	The permutation generation method of W. Myrvold and F. Ruskey \cite{t19};
	\item The numeration systems;
\end{itemize}
to present a new scheme where the secret message embedding capacity is better than the scheme of Aruna Malik and al \cite{t14}.  

\subsection{Permutation Generation Methods}
Permutation is one of the most important combinatorial object in computing, and can be applied in various applications, for example, the scheduling problems. Permutation generation can form the basis of a backtracking program to solve any problem involving reordering a set of items. It is well-known that, for n distinct items, the total number of permutations is n!. 
Permutation generation has a long history. Surveys in the field have been published in 1960 by D.H. Lehmer \cite{t20}. Several authors \cite{t19, t21, t22, t23} have since developed many methods to generate all the possible permutations of n elements.
Also, several works \cite{t13, t24, t25, t26, t27} in steganography taking advantage of permutations have been done. In particular, H. Hioki [26] in 2013, proposed a permutation steganography, which is an effective method for hiding messages provided where the contents of cover objects are not affected by the rearrangement of their elements.
In their paper, W. Myrvold and F. Ruskey \cite{t19} proposed a ranking function for the permutations on n symbols which assigns a unique integer in the range [0, $n$! - 1] to each of the $n$! permutations. Also, they proposed an unranking function for which, given an integer $r$ between 0 and $n$! - 1, the value of the function is the permutation of rank r. Their algorithms is presented below \cite{t19}.

\subsubsection{Unranking function}
First of all, recall that a permutation of order $n$ is an arrangement of $n$ symbols. An array $\pi [0 \cdots n-1]$ is initialized to the identity permutation $\pi[i] = i$, for $i= 0, 1, \cdots n-1$.\\
	
{\bf{Procedure}} $unrank (n, r, \pi)$\cite{t19}\\
$\phantom{sal}$ {\bf{begin}} \\
$\phantom{salut}$ {\bf{if}} $n > 0 $ {\bf{then}} \\
$\phantom{salutsal} swap(\pi[n-1], \pi[r$ mod $n])$; \\
$\phantom{salutsal} unrank(n-1, \left\lfloor r /n\right\rfloor, \pi)$; \\
$\phantom{salut}$ {\bf{end}};\\ 
$\phantom{sal}${\bf{end}};\\

{\bf{Note}}: $swap(a, b)$ exchanges the values of variables $a$ and $b$.
\subsubsection{Ranking function}
To rank, first compute $\pi^{-1}$. This can be done by iterating \\
$\pi^{-1}[\pi[i]] = i$, for $i=0, 1,\cdots, n-1$. \\
In the algorithm below, both $\pi$ and $\pi^{-1}$ are modified.\\

{\bf{function}} $rank (n, \pi, \pi^{-1})$:integer\cite{t19}\\
$\phantom{sal}$ {\bf{begin}} \\
$\phantom{salut}$ {\bf{if}} $n = 1 $ {\bf{then}} return(0) {\bf{end}}; \\
$\phantom{salutsal} s:=\pi[n-1]$ ; \\
$\phantom{salutsal} swap(\pi[n-1], \pi[\pi^{-1}[n-1]])$ ; \\
$\phantom{salutsal} swap(\pi^{-1}[s], \pi^{-1}[n-1])$ ; \\
$\phantom{salutsal}$return$(s+n.rank(n-1, \pi, \pi^{-1}))$ ; \\
$\phantom{sal}${\bf{end}};

\section{Scheme design based on permutation}
In this section, we present the first method of our scheme.

\subsection{Embedding Algorithm}
{\bf{Input}}:\\
$\phantom{saluts}C$: the cover text;\\
$\phantom{saluts}M$: the secret message to embed;\\
$\phantom{saluts}$The key $\pi$: the initial permutation of $n$ colors;\\
$\phantom{saluts} e $: the e-mail address of the receiver\\
{\bf{Output}}:\\
$\phantom{saluts}C'$: the stego-message;\\
{\bf{begin}}:\\
$\phantom{salut}$ 1. Compute $m$, the binary representation of $M$;\\
$\phantom{salut}$ 2. Compute $t= \left\lfloor log_2(n!) \right\rfloor$\\
$\phantom{salut}$ 3. Divide $m$ into $p$ blocks of $t$ bits each, $b_1, b_2, \cdots, b_p$;\\
$\phantom{salut}$ 4. Divide $C$ into $k$ blocks of $n$ characters each $c_1, c_2, \cdots, c_k$;\\
$\phantom{salut}$ 5. For each block $b_i, 1\leq i \leq p$:\\
$\phantom{salutsalut}$ a. compute $Nperm=(b_i)_{10}$, the decimal representation of $b_i$;\\
$\phantom{salutsalut}$ b. compute $\pi' = unrank(n, Nperm, \pi)$, the permutation corresponding to the number $Nperm$. $\pi'$ can be considered as $\pi'(1), \pi'(2), \cdots, \pi'(n)$;\\
$\phantom{salutsalut}$ c. color each character of $c_i$ by the corresponding color given by the permutation $\pi'$ and obtain the string $c'_i$;\\
$\phantom{salutsalut}$ d. compute $C' \leftarrow C'||c'_i$; where a$||$b is the concatenation of a and b.\\
$\phantom{salut}$ 6. If the next character is EOF (End of File)  then\\
$\phantom{salutsalu}$ {\bf{begin}}\\
$\phantom{salutsalut}$ a.	Use e to send C' by mail to the receiver;\\
$\phantom{salutsalu}$ {\bf{end}}\\
$\phantom{salutsal}$ {\bf{Else}}\\
$\phantom{salutsalu}$ {\bf{begin}}\\
$\phantom{salutsalut}$ a. Colour the next character with a color different of permutation colors. This color is shared by the sender and the receiver. However, this color will not be very distant from the others.\\
$\phantom{salutsalut}$ b. Randomly color the rest of characters of C by the colors of colors table, until the EOF character is obtained;\\
$\phantom{salutsalut}$ c.	Use e to send C' by mail to the receiver;\\
$\phantom{salutsalu}$ {\bf{end}};\\
$\phantom{sal}${\bf{end}};

\subsection{Retrieval Algorithm}
{\bf{Input}}:\\
$\phantom{saluts}C'$: the stego-text;\\
$\phantom{saluts}$The key $\pi$: the initial permutation of $n$ colors;\\
{\bf{Output}}:\\
$\phantom{saluts}M$: the secret message;\\
{\bf{begin}}:\\
$\phantom{salut}$ 1. Retrieve all characters coloured by the permutation colors, until a color different from the colors in the colors table, or the EOF character is obtained. Lets call them $C''$;\\
$\phantom{salut}$ 2. Divide $C''$ into $p$ blocks of $n$ characters each $c_1, c_2, \cdots, c_p$;\\
$\phantom{salut}$ 3. For each block $c_k, 1\leq k \leq p$:\\
$\phantom{salutsalut}$ a. use the color order of characters to compute the relative permutation, that we call $\pi' $. $\pi'$ can be considered as $\pi'(1), \pi'(2), \cdots, \pi'(n)$;\\
$\phantom{salutsalut}$ b. compute the number $Nperm = rank(n, \pi', \pi'^{-1})$. \\
$\phantom{salutsalut}$ c. compute $m'=(Nperm)_2$, the binary representation of $Nperm$;\\
$\phantom{salutsalut}$ d. compute $M \leftarrow M||m'$;\\
$\phantom{sal}${\bf{end}};

\subsection{Experimentation}

\begin{figure}[h]
\centering
{\includegraphics[width=6 cm, height= 3.9cm]{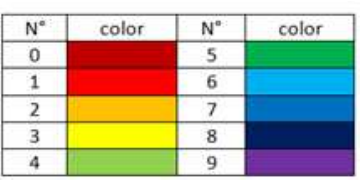}
 \caption{\rm The table of 10 colors}.
\label{aa}}
\end{figure}

\begin{figure}[h]
\centering
{\includegraphics[width=6 cm, height= 3.9cm]{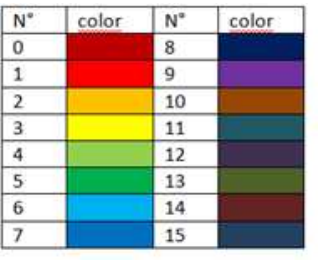}
 \caption{\rm The table of 16 colors}.
\label{a}}
\end{figure}

\begin{figure}[h]
\centering
{\includegraphics[width=8 cm, height= 4.5cm]{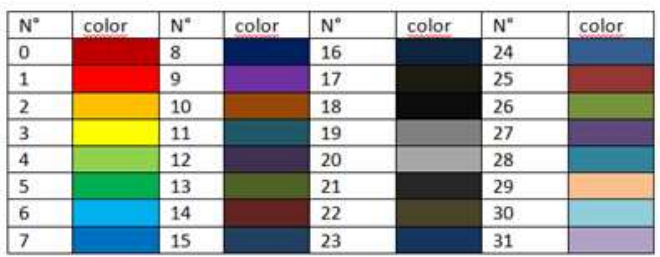}
 \caption{\rm The table of 32 colors}.
\label{b}}
\end{figure}

\begin{figure}[h]
\centering
{\includegraphics[width=9 cm, height= 5.3cm]{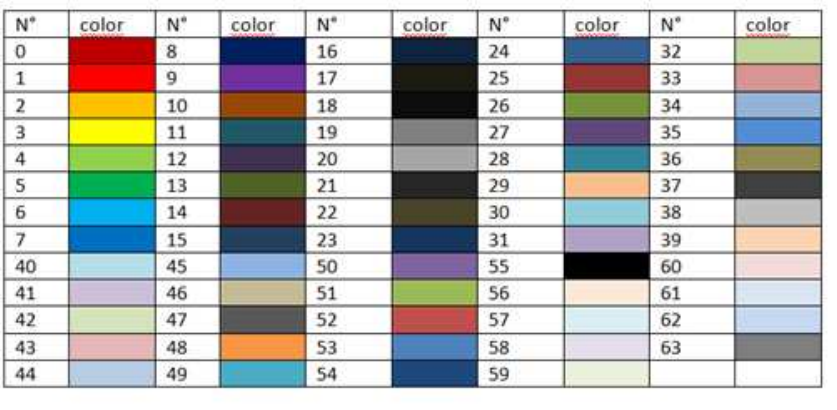}
 \caption{\rm The table of 64 colors}.
\label{c}}
\end{figure}

In this subsection, we present some experimentations of this method. First, we propose a theoretical estimation of our embedding capacity for $n$ colors. Secondly, we present a practical experimentation in the case of 10, 16, 32 and 64 colors, based on example 1 and figure 5 of \cite{t14}. These colors are given in figure \ref{aa}, figure \ref{a}, figure \ref{b} and figure \ref{c}.\\

In order to present our embedding capacity, we use as cover text and secret message those of example 1 and figure 5 of Aruna Malik and al \cite{t14}.

\subsubsection{Theoretical estimations}
The table \ref{ttt} presents the embedding capacity of our scheme for some different values of n: 10, 16, 20 32,60, 64. This theoretical estimation is based on our embedding algorithm. \\
\indent More generally, in a set of n colors, the number of permutation of n distinct colors is n !. According to the stirling formula \cite{t28} we have:
 \begin{equation}
n! \sim \left(\frac{n}{e}\right)^n \times \sqrt{2\pi n}
 \end{equation}

Where $\pi=3.14$ is the area of the circle with unit radius, $e=2.718$ is the base of the natural logarithm, and $\sim$ means approximate equality.

we know that: 
\begin{equation*}
n=2^{log_2(n)}
 \end{equation*}

By replacing the value of n in equation 1 we have:
\begin{align*}
n! & \sim \left( \frac{2^{log_2(n)}}{2^{1.442695}} \right)^n \times  \sqrt{2\pi n} \\
   & \sim \left( 2^{log_2(n)-1.442695}\right)^n \times \sqrt{2\pi n} \\
   & \sim \left( 2^{nlog_2(n)-1.442695n}\right) \times \sqrt{2\pi n} \\
   & \sim \left( 2^{nlog_2(n)-1.442695n}\right) \times 2^{log_2(\sqrt{2\pi n})} \\
   & \sim \left( 2^{nlog_2(n)-1.442695n}\right) \times 2^{\frac{1}{2}log_2(2\pi n)} \\
   & \sim \left( 2^{nlog_2(n)-1.442695n+ {\frac{1}{2}log_2(2\pi n)}}\right)  \\   
 \end{align*}

 \textbf{Proposition :} the embedding capacity (E) using n colors to hide a secret is : 
 
 \[E = \frac{M \times 100}{n \times 8}\]\\
where $   M = {n(log_2(n)-1.442695)+ {\frac{1}{2}log_2(2\pi n)}} $, and $n$ the number of colors.

$\blacksquare$

\begin{table}
\begin{tabular}{|c|c|c|c|}
\hline
n 	& 	 M= $ \left\lfloor log(n!) \right\rfloor $ & P=M/8	&	100*(P/n),(embedding capacity)\\
\hline
10 	& 	21 			&   2.6 	& 26.25\% \\
\hline
16 	& 	44 			&   5.5 	& 34.37\% \\
\hline
20 	& 	61 			&   7.6 	& 38\% \\
\hline
32 	& 117 			&   14.6 	& 45.63\% \\
\hline
60 	& 	272 		&   34 	& 56.67\% \\
\hline
64 	& 	295 		&   36.9 	& 57.66\% \\
\hline
\end{tabular}
  \caption{Theoretical estimations of the proposed scheme}
  \label{ttt}
\end{table}

{\bf{Remark:}} As far as the space characters of the stego-text are not coloured, the embedding capacity can decrease in the experimentations.\\

\subsubsection{Experimentation 1}
Here, the secret message is :
{\bf{underlying physiological mechanisms}}\\
and the cover text is:\\
{\bf{Only boats catch connotes of the islands sober wines only ships wrap the slips on the cleats of twining lines only flags flap in tags with color that assigns only passage on vessels}}\\

Here we present the embedding process.
\begin{enumerate}
	\item We compute the binary representation of the secret and obtain the following result:\\
	
	01110101   01101110  01100100  01100101  01110010  01101100  01111001  01101001  01101110   01100111          00100000
01110000   01101000  01111001  01110011  01101001  01101111  01101100  01101111  01100111   01101001  01100011  01100001  01101100      00100000
01101101  01100101  01100011  01101000  01100001  01101110   01101001    01110011  01101101  01110011

  \item We compute $t= \left\lfloor log_2(10!) \right\rfloor = 21 $ ; 
  
  \item The binary secret is then divided into blocks of 21 bits each. For instance, the first block $b_1 $ = 011101010110111001100 and the second block $b_2$ =   100011001010111001001, ...\\

  \item We divide the cover text into blocks of 10 characters. For instance, the first block $c_1=$ {\bf{Only boats c}}, the second block $c_2= $ {\bf{atch connot}}, ...;
  
  \item We color the cover text:
  
\begin{itemize}
	\item For the block $b_1 $ = 011101010110111001100, Nperm = 961996, its decimal representation.\\
	\item The permutation relative to 961996 is given by $\pi'$ = unrank(10, 961996, $\pi )$ = 3  8  5  2  1  4  9  0  7  6, $\pi$ = 0 1 2 3 4 5 6 7 8 9, with the corresponding colors given by figure \ref{aa}. \\
	
	\item The block $c_1=$ {\bf{Only boats c}} is coloured relatively to the permutation $\pi'$. We then obtain the color text given in figure \ref{d1}

	\begin{figure}[ht]
	\centering
{\includegraphics[width=4 cm, height= 1.5cm]{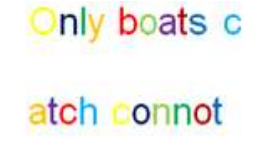}
 \caption{\rm The Stego-Text}.
\label{d1}}
\end{figure}

\end{itemize}

\begin{itemize}
	\item For the block $b_2 $ = 100011001010111001001, Nperm = 1152457, its decimal representation.\\
	\item The permutation relative to 1152457 is given by $\pi'$ = unrank(10, 1152457, $\pi )$ = 2  9  1  6  3  8  4  5  0  7, $\pi$ = 0 1 2 3 4 5 6 7 8 9, with the corresponding colors given by figure \ref{aa}. \\
	
	\item The block $c_2=$ {\bf{atch connot}} is coloured relatively to the permutation $\pi'$. We then obtain the color text given in figure \ref{d1}
	
\end{itemize}
 
 \item The process is the same, and finally we obtain the stego-text given by the figure \ref{fig10}. That stego-text is then send by mail to the receiver.

\end{enumerate}

 \begin{figure}[ht]
{\includegraphics[width=12 cm, height= 2.0cm]{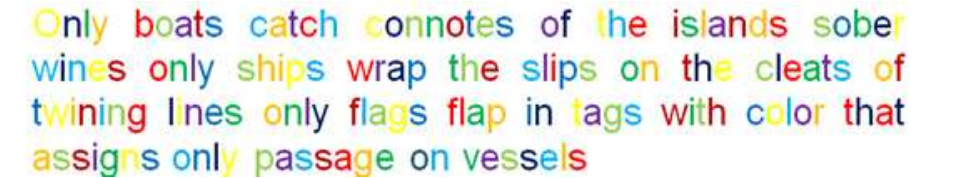}
 \caption{\rm The Stego-Text}.
\label{fig10}}
\end{figure}

With this example :
\begin{itemize}
	\item in the case of 10 colors, the embedding capacity is 20.58 \%;
	\item with 16 colors, the embedding capacity is 25.5 \%;
	\item With 32 colors, the embedding capacity is 29.5 \%;
	\item with 64 colors, the embedding capacity is 45.45 \%.
\end{itemize}

\subsubsection{Experimentation 2}

In the example of figure 5 \cite{t14}, the secret message is :
{\bf{behind using a cover text is to hide the presence  of secret messages the presence of embedded messages in the resulting stego-text cannot be easily discovered by anyone except the intended recipient.}}\\

and the cover-text is:

{\bf{in the research area of text steganography, algorithms based on font format have advantages of great capacity, good imperceptibility and wide application range. However, little work on steganalysis for such algorithms has been reported in the literature. based on the fact that the statistic features of font format will be changed after using font-format-based steganographic algorithms, we present a novel support vector machine-based steganalysis algorithm to detect whether hidden information exists or not. this algorithm can not only effectively detect the existence of hidden information, but also estimate the hidden information length according to variations of font attribute value. as shown by experimental results, the detection accuracy of our algorithm reaches as high as 99.3 \% when the hidden information length is at least 16 bits. Our scheme present experimentation based on different colors number.}}

\begin{figure}[ht]
{\includegraphics[width=14 cm, height= 4.5cm]{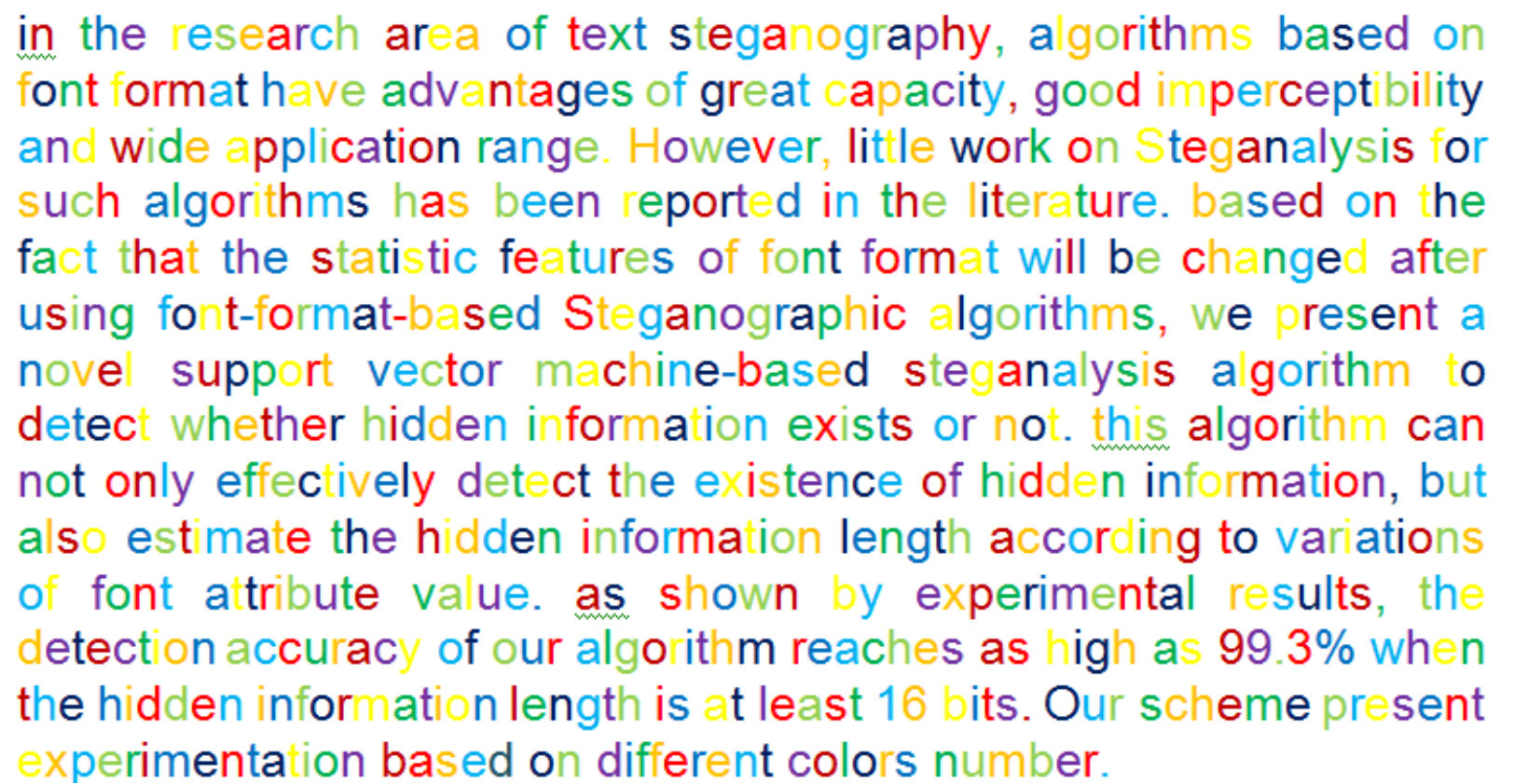}
 \caption{\rm The Stego-Text}.
\label{e}}
\end{figure}

 In the  case of 10 colors, We apply our embedding algorithm and obtain the following stego-text, given by figure \ref{e}.

With this example:
\begin{itemize}
	\item In the case of 10 colors, the embedding capacity is 22.32 \%;
	\item With 16 colors, the embedding capacity is 29.64 \%;
	\item With 32 colors, the embedding capacity is 38 \%;
	\item With 64 colors, the embedding capacity is 44 \%.
	
\end{itemize}

\section{Scheme design based on numeration systems}
In this new approach, we improve the method of the first scheme with the assertion that each color can be repeated as many times on some positions of a given group of characters. Unlike the previous scheme in which each color could only appear once in a group of precise characters. 

\subsection{The Scheme description}
we give a brief description of how this new scheme works by following these steps:
\begin{enumerate}
\item Choose a base $B$ such that $2 \leq B \leq 2^{24} $. where $2^{24}$ is the number of existing colors ;
\item choose B colors from the set of $2^{24}$ colors number from 0 to $B -1$;
\item convert the secret m to base B such that : $m=(m_{q-1} ...  m_1m_0)_B$, \\
where  $0 \leq m_i \leq B-1$;

\item We assume that the number of characters of the covert text is : $n$ and $q \leq n$; 

\item For $i=0$ to $q-1$  do \\         
       The character $c_{i}$ is coloured with the color relative to $m_{i}$
\item The text coloured is then send to the receiver.

\end{enumerate}

The reverse procedure consists to extract the secret conceal in the colors distribution. These steps must be performed by the receiver of the stego-text : 
\begin{enumerate}
\item Take the text with the first $q$ characters which has been coloured;
\item For $i=0$ to $q-1$ do \\
      
       Find the color number $z_{i}$ associated to the character $c_{i}$  
      by using the reference color table shared between the sender and 
      the receiver;
 \item Convert $z=(z_{q-1} ...  z_1z_0)_B$ to binary and get the secret message.
\end{enumerate}

\subsection{Embedding Algorithm}

\textbf{Input}

C: the cover text;

M: the secret message to embed;

$\beta$ : The base;

T : a table of $\beta$ color;

e : the e-mail address of the receiver;
\\
\textbf{Output}

C': the stego-message;
\\
\textbf{Begin}

\begin{enumerate}
\item Convert the secret M to base B such that : $m=(m_{n-1} ...  m_1m_0)_B$, \\
where  $0 \leq m_i \leq B-1$;
\item For $i=n-1$ to $0$ do 
\begin{enumerate}
      \item Find in the color table, the color $a_i$ associated  to the value  $m_i$;
      \item Coloured the character $c_i$ of  $C$ with the color $a_i$ and obtain $c'_i$ ;
      \item Compute $C' \longleftarrow C' \mid \mid c'_i$; where $a\mid\mid b$ is the concatenation of a and b.
\end{enumerate}     

\item If the next character is not EOF (End of File) then
\begin{enumerate}
\item Colour the next character with a color different from the colors table T. This color is shared by the sender and the receiver. However, this color will not be very distant from the others;
\item Randomly color the rest of characters of C by the colors from the colors table, until obtain the EOF character;
\item Compute $C' \longleftarrow C' \mid \mid c'_j$ : $n \leq j \leq m$, where $m$ is the position of the last character of $C$;
\end{enumerate}
\item Use e to send $C'$ by mail to the receiver;
\end{enumerate}
\textbf{End}

\subsection{Retrieving Algorithm}
\textbf{Input}

C': the stego-text;

T : a table of $\beta$ color;

$\beta$ : The base;
\\
\textbf{Output}

M: the secret message;
\\
\textbf{Begin}
\begin{enumerate}
\item Retrieve all characters coloured with the table colors, until obtain a color different from those of the colors table, or obtain the EOF character. Lets call them $C''$ and $\mid C'' \mid = n$; ($C'' = c_{n-1}c_{n-2}...c_1c_0$);

\item For $i=n-1$ to $0$ do
\begin{enumerate}
      \item get the color $a_i$ associated to the color of the character  $c_i$ of $C''$;
      \item Find in the color table, the value $m_i$  associated to the color $a_i$    ;
       \item compute $M \longleftarrow M \mid \mid m_i $;
      \item Compute $M_2$, the binary representation of the secret $M$;
\end{enumerate}  

\end{enumerate}

\textbf{End}

\subsection{Experimentation}
This subsection presents some experimentations for this method. We first propose a theoretical estimation of our embedding capacity for $B$ colors. Secondly, we present a practical experimentation in the case of 10, 16 and 32 colors, based on example 1 and figure 5 of \cite{t14}. These colors are given in figure \ref{aa}, figure \ref{a} and figure \ref{b}.\\

\subsubsection{Theoretical Estimation}

We want to color a block of text with $\eta$ characters. Each character is coloured with a single color. The number of colors used is $B$. Knowing that a color can appear as many times on some positions, the total number of colouring possibilities for each character is : $B$. For the $\eta$ characters, the total number of colouring possibilities is : $B^{\eta}$.

The number of bits used to color the $\eta$ characters is : $log_2(B^{\eta})$

The embedding capacity \cite{t14,t18} is define as the ratio of the secret bits message by the stego cover  bits : 
\begin{align}
Capacity & =  \frac{Bits \; of \; secret \;  message}{Bits \; of \; stego \; cover} \\ 
  Capacity &= \frac{log_2(B^{\eta})}{\eta \times 8}    \\
  Capacity &= \frac{log_2(B)}{ 8}   
\end{align}

The following table gives a theoretical estimate of the capacity as a function of the base $B$ used:

 \begin{table}[ht]
 \centering
\caption{\textbf{Embedding capacity estimation as a function of $B$}}
\begin{tabular}{cc}
\\

\hline
$B$ 	& 	Capacity $\times 100$	 \\
\hline
2	& 	 12.5\% \\
 
4	& 	 25\% \\
 
8	& 	 37.5\% \\

10	& 	 41.5\% \\

16	& 	 50\% \\

32	& 	 62.5\% \\
 
64	& 	 75\% \\
\hline 
\end{tabular}
\end{table}

\subsubsection{Experimentation 1}

\begin{figure}[ht]
{\includegraphics[width=12 cm, height= 2.0cm]{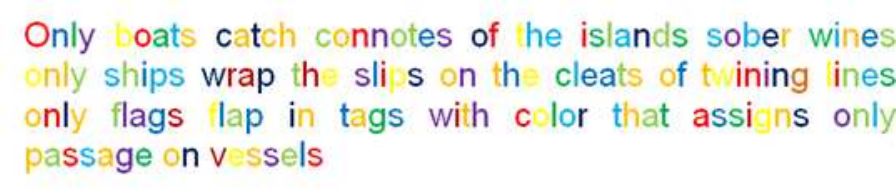}
 \caption{\rm The Stego-Text for a table of 10 colors}.
\label{fig14}}
\end{figure}

This experimentation is based on example 1 of \cite{t14}, where the number of color $B$ is equal to 10. The figure \ref{fig14} presents the results of the embedding process based on this second method for 10 colors.

With this example :
\begin{itemize}
	\item in the case of 10 colors, the embedding capacity is 34.31 \%;
	\item with 16 colors, the embedding capacity is 41.17  \%;
	\item With 32 colors, the embedding capacity is 52.23  \%;
\end{itemize}
\subsubsection{Experimentation 2}

\begin{figure}[ht]
{\includegraphics[width=14 cm, height= 4.5cm]{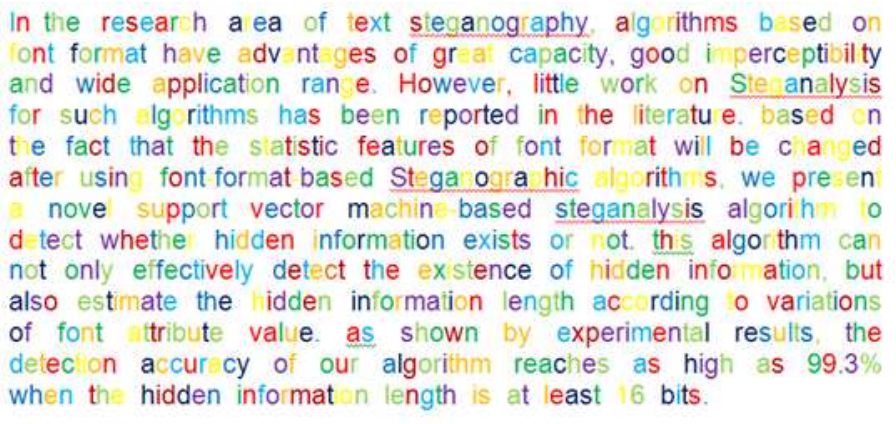}
 \caption{\rm The Stego-Text for a table of 10 colors}.
\label{fig17}}
\end{figure}

The figure \ref{fig17} presents the results of our embedding process for 10 colors, based on the example gives by figure 5 of \cite{t14}. 

With this example :
\begin{itemize}
	\item in the case of 10 colors, the embedding capacity is 35.29 \%;
	\item with 16 colors, the embedding capacity is 42.85  \%;
	\item With 32 colors, the embedding capacity is 53.22  \%;
\end{itemize}

The table \ref{15} recapitulates the embedding capacity of our schemes in comparison with the scheme of Aruna and al \cite{t14}, in the case of 10 colors.

\begin{table}
\begin{center}
\begin{tabular}{|l|r|r|r|}
\hline
\phantom                       & {\bf{ First Method}} & {\bf{Second Method}} & The scheme of Aruna and al \cite{t14} \\ \hline
example 1 \cite{t14}           & {\bf{ 20.58 \% }}  & {\bf{34.31 \%}} & 6.03 \%  \\ \hline
example of figure 5 \cite{t14} & {\bf{ 22.32 \% }}  & {\bf{35.29 \%}} & 13.43\% \\ \hline
\end{tabular}
\caption{Comparison between our scheme and the scheme of Aruna \cite{t14}, in terms of embedding capacity, for 10 colors}
\label{15}
\end{center}
\end{table}

\section{Conclusion}
In this paper, two text steganographic schemes based on color
coding have been proposed.  The first based on permutation and the second based on 
numeration systems. Given a secret message and a cover text, the proposed schemes embed the secret 
message in the cover text by making it coloured. Using 32 colors, the first scheme achieves a theoretical and 
practical embedding capacity of 45.63 \%  and 38 \% respectively. While with the second scheme the theoretical and 
practical embedding capacity are 62.5\% and 53.22\% respectively with the same number of colors. These two high
capacity text steganographic scheme significantly improve the existing work of Aruna and al.

\section{Acknowledgments}
This work was supported by UMMISCO, CETIC and the University of Yaounde 1.






\end{document}